\documentclass{llncs}

\usepackage{graphicx}
\usepackage{color}
\usepackage{amssymb}
\usepackage{epsfig}
\usepackage{graphicx}
\usepackage{epsfig}
\usepackage{psfrag}

\usepackage{times}

\usepackage{amsmath}
\usepackage{amssymb}
\usepackage{algorithm2e}
\makeatletter
\renewcommand{\@algocf@capt@plain}{above}
\makeatother

\begin{document}

\title{Classification of lung nodules in CT images based on Wasserstein distance in differential geometry}

\author{Min Zhang \inst{1} \and Chengfen Wen \inst{2} \and Qianli Ma \inst{3,4} \and  Huai Chen \inst{5}\and Deruo Liu \inst{4} \and  Jie He \inst{3}\and  Xianfeng Gu \inst{2} \and  Xiaoyin Xu\inst{1} }

\institute{Department of Radiology, Brigham and Women's Hospital, Harvard Medical School, Boston, MA 02115, US\\
           \and Department of Computer Sciences, Stony Brook University, Stony Brook, NY 11794, US\\
           \and Department of General Thoracic Surgery, Cancer Hospital, Chinese Academy of Medical Science, Beijing 100021, China\\
           \and Department of Thoracic Surgery, China-Japan Friendship Hospital, Beijing 100029, China \\
           \and Department of Radiology, First Affiliated Hospital of Guangzhou Medical University, Guangzhou 510120, China
           }
           \author{Min Zhang \inst{1} \and Chengfen Wen \inst{2} \and Qianli Ma \inst{3,4} \and  Huai Chen \inst{5}\and Deruo Liu \inst{4}\and  Jie He \inst{3}\and  Xianfeng Gu \inst{2} \and  Xiaoyin Xu\inst{1} }

\maketitle

\begin{abstract}

Lung nodules are commonly detected in screening for patients with a risk for lung cancer.
Though the status of large nodules can be easily diagnosed by fine needle biopsy or bronchoscopy,
small nodules are often difficult to classify on computed tomography (CT).
Recent works have shown that shape analysis of lung nodules can be used to differentiate benign lesions
from malignant ones, though existing methods are limited in their sensitivity and specificity.
In this work we introduced a new 3D shape analysis within the framework of differential geometry to calculate the
Wasserstein distance between benign and malignant lung nodules to derive an accurate classification scheme.
The Wasserstein distance between the nodules is calculated based on our new spherical optimal mass transport,
this new algorithm works directly on sphere by using spherical metric, which is much more accurate and efficient than previous methods.
In the process of deformation, the area-distortion factor gives a probability measure on the unit sphere, which forms the Wasserstein space.
From known cases of benign and malignant lung nodules, we can calculate a unique optimal mass transport map between their
correspondingly deformed Wasserstein spaces. This transportation cost defines the Wasserstein distance between them
and can be used to classify new lung nodules into either the benign or malignant class.
To the best of our knowledge, this is the first work that utilizes Wasserstein distance for lung nodule classification.
The advantages of Wasserstein distance are it is invariant under rigid motions and scalings,
thus it intrinsically measures shape distance even when the underlying shapes are of high complexity,
making it well suited to classify lung nodules as they have different sizes, orientations, and appearances.

\end{abstract}
\vspace{-10mm}
\section{Introduction}\label{sec:Introduction}
\vspace{-4mm}
Lung nodules are commonly detected in screening for patients with a risk for lung cancer. Though the status of large nodules can be easily diagnosed by fine needle biopsy or bronchoscopy, small nodules are very difficult to assess,especially if they are located deep in the tissue or away from the large airways. Most of the small nodules can only be diagnosed in CT images as many visual features are extracted from the CT scan data for classifying a specific nodule to be benign or malignant. Some recent studies show that the shape of lung nodules is an important feature for distinguishing benign lesions from malignant ones. Clinicians usually classify nodules, based on their morphologic characteristics, and make predictions about  the "level of suspicion" \cite{Thorac2015}. The above manual analysis, however, is an error-prone and time consuming process. In this work, we introduced a new approach to classify detected lung nodules based on their shapes by using Wasserstein distance within the framework of differential geometry.

Wasserstein Distance has been widely studied for shape analysis, and it has been also used for medical image analysis~\cite{PAMISu2015}. In our work, the 3D shape of a lung nodule are conformally mapped to the unit sphere and the conformal map will distort the surface area. The area-distortion factor gives a probability measure on the unit sphere as measured from the Wasserstein space. Given any two probability measures, there is unique optimal mass transport map between them and the Wasserstein distance defined as the transportation cost between them. It has significant power to measure the intrinsic differences between shapes and which can be used for shape classification.

\begin{figure}[t]
\vspace{-3mm}
\begin{center}
\vspace{-3mm}
\begin{tabular}{ccc}
\includegraphics[width=0.3\textwidth]{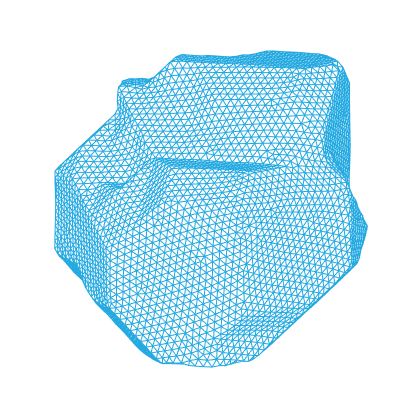}&
\includegraphics[width=0.3\textwidth]{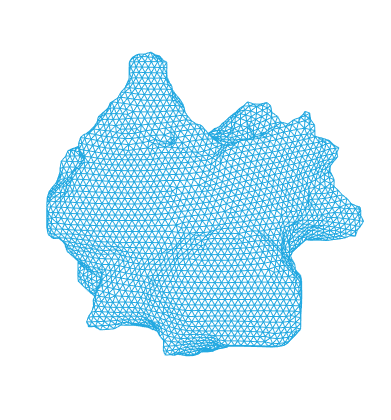}&
\includegraphics[width=0.35\textwidth]{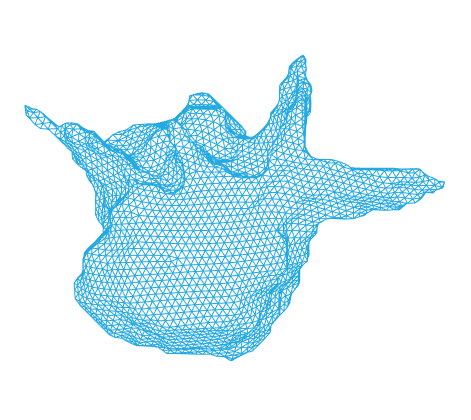}
\\
(a) spherical& (b)lobulation &(c) spiculation\\

\end{tabular}
\end{center}
\caption{Lung nodules with different morphology
}
\label{fig:original}
\vspace{-3mm}
\end{figure}

The ultimate goal of the classification is to differentiate the malignant nodules from benign ones, so we classify the nodules to two different category malignant and benign base on their Wasserstein Distance between the lung nodules with different morphology (Figure ~\ref{fig:original}).

To the best of our knowledge, this is the first work that utilizes the Wasserstein distance for lung nodule classification. Wasserstein distances are invariant under rigid motions and scalings, thus it intrinsically measures shape distance even when the underlying shapes are of high complexity, making it well suited to classify lung nodules as they have different sizes, orientations, and appearances. We present a more efficient Wasserstein distance calculation algorithm under spherical metric.
\vspace{-3mm}
\begin{figure}
\begin{center}
\begin{tabular}{ccc}
\includegraphics[width=0.33\textwidth]{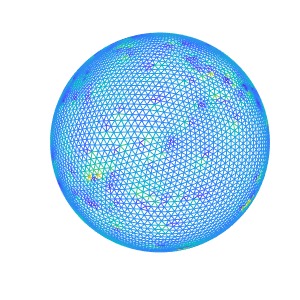}&
\includegraphics[width=0.33\textwidth]{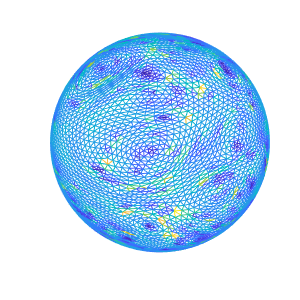}&
\includegraphics[width=0.33\textwidth]{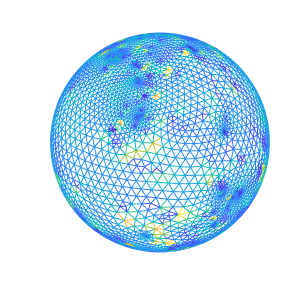}
\\
(a) source & (b) result & (c) target

\end{tabular}
\end{center}
\caption{Optimal Mass Transport map for Sphere}
\label{fig:OMT}
\end{figure}

\vspace{-3mm}
\section{Related Work}\label{sec:RelatedWork}
\vspace{-3mm}

Pulmonary nodules are often detected in high-risk population of lung cancer~\cite{Nantional2003}.
Over the years, there has been a large body of work for lung nodules detection and diagnosis~\cite{JMI2016,LEE201043,Way2009ComputeraidedDO}.
Studies have shown that morphologic features play an important role for discriminating malignant nodules from benign ones
on computed tomography (CT)~\cite{Radiology1993}.
Also studies have shown that quantitative analysis of the morphologic futures of lung nodules will improve radiologists'
classification of malignant and benign nodules~\cite{AJR2004,Radiology2006}.

Most of current approaches for classifying lung nodule are based on 2D features.
Mori et al.~\cite{JCAT2005} evaluate temporal changes as a feature of benign and malignant nodules based in a curvedness index
in combination with dynamic contrast-enhanced CT. Gurney et al.~\cite{Radiology1995} designed CAD systems based on neural network for lung nodule classification.
Furuya et al.~\cite{Acta1999} analyzed margin characteristic of 193 pulmonary nodules and classified the lung nodules based on the analysis.
They found that $97\%$ spiculated and $82\%$ lobulated ones were malignant, and $66\%$ round were proved to be benign.
Several groups developed the approaches based on 3D lung nodules models.
For example, Kawata et al.~\cite{CIP2001} classified the lung nodules based on the 3D
curvatures and the relationship of the nodules to their surrounding features.
Ei-Baz et al presented~\cite{Ipmi2011} a 3D shape analysis method based on a shape index called spherical harmonics
was introduced and applied for lung nodules' classification.
Kurtek et al.~\cite{KurtekPAMI2012} provided a Riemannian framework to compare and match the 3D shapes by computing geodesic paths.
Mahmoudi et al.~\cite{Mahmoudi:2009:GM} compute the histogram of pairwise diffusion distances between all points to describe the shapes.
Jermyn et al.~\cite{Ian:ECCV:2012} compare and analyze the shape based on the definition of general elastic metric on the space of parameter domains.

Our 3D shape analysis approach is based on the spherical Wasserstein distance and we show that it has unique advantages over
above mentioned 2D and 3D approaches. The existing 2D methods did not consider the 3D features of the lung nodules which are
important to describe their pathological characteristics.
The existing 3D methods have their own limitations as they typically can not quantitative measure the differences between two complex shapes, as we encounter
in lung nodules.

Comparing to the conventional methods,our shape classification method only depends on Riemannian metrics and is invariant under
rigid motions and scalings which is more effective and efficient for shape classification.
\vspace{-3mm}
\section{Methods}
\vspace{-3mm}
This section gives the brief introduction to the theoretic background and the computation method.
\subsection{Theoretic Background}
Here we only introduce the most fundamental concepts and theorems, for detailed treatments, we refer readers to \cite{gu:08:ComputationalConformalGeometry} for conformal geometry,
and ~\cite{Kantorovich48}, \cite{Brenier} for optimal mass transportation.

\subsubsection{Optimal Mass Transport}

Optimal mass transportation problem was first raised by Monge \cite{Monge} in the 18th century. In the following, we formulate the problem
in general Riemannian manifold setting.

Suppose $(M,\mathbf{g})$ is a Riemannian manifold with a Riemannian metric $\mathbf{g}$, let $\mu$ and $\nu$ be two probability measures on $M$, which have the same total mass $\int_M d\mu = \int_M d\nu$. A map $T:M\to M$ is \emph{measure preserving}, if
for any measurable set $B\subset M$, $\int_{T^{-1}(B)} d\mu = \int_B d\nu$. Namely, $T$ pushes $\mu$ forward to $\nu$, denoted as $T_\#\mu=\nu$. The optimal mass transportation problem is formulated as follows:
\begin{problem}[Optimal Mass Transport] Given a transportation cost function $c: M\times M\to \mathbb{R}$,
find the measure preserving map $T:M\to M$ that minimizes the total
transportation cost
\begin{equation}
    \mathcal{C}(T):=\int_M c(x,T(x)) d\mu(x).
    \label{eqn:transportation_cost}
\end{equation}
\end{problem}
In our current work, the cost function is the squared geodesic distance, $c(x,y) = d_{\mathbf{g}}^2(x,y)$.

In the 1940s, Kantorovich solved the optimal transportation problem by relaxing the transportation maps to transportation plans, and introduced the linear programming method~\cite{Kantorovich48}.

\begin{theorem}[Kantorovich] Suppose $(M,\mathbf{g})$ is a Riemannian manifold, probability measures $\mu$ and $\nu$ have the same total mass, $\mu$ is absolutely continuous, $\nu$ has finite second moment, the cost function is the squared geodesic distance, then the optimal mass transportation map exists and is unique.
\end{theorem}

\subsubsection{Wasserstein Metric Space}
All the probability measures on the Riemannian manifold form the Wasserstein distance, the optimal transportation cost defines the metric of this space.
\begin{definition}[Wasserstein Space]
For $p\ge 1$, let $\mathcal{P}_p(M)$ denote the space of all
probability measures on $M$ with finite $p^{th}$ moment, for
some $x_0\in M$, $\int_M d(x,x_0)^p d\mu(x) < +\infty,$ where $d$ is
the geodesic distance induced by $\mathbf{g}$.
\end{definition}

Given two probabilities $\mu$ and $\nu$ in $\mathcal{P}_p$,
the Wasserstein distance between them is defined as the
transportation cost induced by the optimal transportation map
$T:M\to M$,
\[
    W_p(\mu,\nu) := \inf_{T_\#\mu = \nu} \left( \int_M d_{\mathbf{g}}^p(x,T(x))
    d\mu(x) \right)^{\frac{1}{p}}.
\]
 The following theorem plays a fundamental role for
the current work
\begin{theorem}\cite{Villani:OMT}
The Wasserstein distance $W_p$ is a Riemannian metric of the
Wasserstein space $\mathcal{P}_p(M)$.
\end{theorem}
\subsubsection{Discrete Optimal Mass Transport}

Suppose the target measure $\nu$ is a Dirac measure, namely, its support is a discrete point set, $P=\{p_1,p_2,\cdots,p_k\}$. Each point $p_i$ is with measure $\nu(p_i) = \nu_i$. Then the optimal mass transportation map is given by a geodesic power diagram.


\begin{definition}[Geodesic Power Diagram] Given a point set $P=\{p_1,p_2,\cdots,p_k\}$ and the weight $\mathbf{h}=\{h_1,h_2,\cdots,h_k\}$, the geodesic power diagram induced by $(P,\mathbf{h})$ is a cell decomposition of the manifold $(M,\mathbf{g})$, such that the cell associated with $p_i$ is given by
\[
    W_i:=\{q\in M| d_\mathbf{g}^2(p_i,q) + h_i \le d_\mathbf{g}^2(p_j,q) + h_j \}.
\]
\end{definition}
The following theorem lays down the foundation of our algorithm.
\begin{theorem}[Discrete Optimal Mass Transport] Given a Riemannian manifold $(M,\mathbf{g})$, two probability measures $\mu$ and $\nu$ are of the same total mass. $\nu$ is a Dirac measure, with discrete point set support $P=\{p_1,p_2,\cdots,p_k\}$, $\nu(p_i)=\nu_i$. There exists a weight $\mathbf{h}=\{h_1,h_2,\cdots,h_k\}$, unique up to a constant, the geodesic power diagram induced by $(P,\mathbf{h})$ gives the optimal mass transportation map,
\[
    T: W_i \to p_i, i = 1, 2,\cdots, k,
\]
furthermore $ \int_{W_i} d\mu = \nu_i, \forall i$.
\end{theorem}

\subsubsection{Discrete Spherical Optimal Mass transport}

Let $\mathbb{S}^2$ be the unit sphere. A \emph{great circle} is a circle which intersects $\mathbb{S}^{2}$ with a plane passing through the center. The \emph{geodesic} between two points $A,B$ on $\mathbb{S}^{2}$ is a portion of the great circle that passes through these two points. The geodesic distance $d(p,q) = \cos^{-1}(p\cdot q)$, where \textbf{$\cdot$} is the ordinary dot product.

Suppose that we are given a set of circles $G=\{c_{1},c_{2},\cdots,c_{n}\}$ on $\mathbb{S}^{2}$. Laguerre proximity~\cite{sugihara2002laguerre} is defined as
$
d_{L}(p,c_i)=\frac{\cos d(p,p_i)}{\cos r_i}.
$
The region which consists of the points that are closer to $c_i$ than any other circles in $G$ is defined by
$
R(G;c_i)=\{p\in\mathbb{S}^{2}\mid d_{L}(p,c_i) < d_{L}(p,c_j), j\neq i\}.
$
The partition $\mathcal{V}=\{R(G;c_i),i=1,2,\cdots,n\}$ of $\mathbb{S}^{2}$ by these regions gives the \emph{spherical power diagram} for $G$. The Poincare dual of the spherical power diagram gives the spherical Delaunay triangulation. All the edges of the triangulation and the spherical power diagram are geodesic. Moreover, every edge of power diagram is perpendicular to its dual edge in triangle mesh.

The measure for each cell $R(G,c_i)$ is denoted as $A_i$, 
\[
	A_i = \int_{R(G,c_i)} d\mu,
\]
The energy defined on the radii of $c_i$'s is given by
\[
E(\mathbf{r}) = \int^{\mathbf{r}} \sum_{i=1}^k (A_i-\nu_i) dr_i. 
\]
The gradient of $E$ equals to the difference between the current measure and target measure. For the Hessian of energy $E$, the symmetry can be obtained by direct calculation. Since $\frac{\partial{E}}{\partial{r_i}} = A_i - \nu_i$, $\frac{\partial^2{E}}{\partial{r_i}^2} = \frac{\partial{A_i}}{\partial{r_i}}$. Since $\sum_i{A_i} = \sum_i{\int_{W_i}{\mu}} = \int_{S_1}{\mu} = 4\pi$, by symmetry we have
$$
\frac{\partial^2{E}}{\partial{r_i}^2}
= \frac{\partial{A_i}}{\partial{r_i}}
= -\sum_{j\neq i}{\frac{\partial{A_j}}{\partial{r_i}}}
= -\sum_{j\neq i}{\frac{\partial{A_i}}{\partial{r_j}}}
= -\sum_{j\neq i}{\frac{\partial^2{E}}{\partial{r_i}\partial{r_j}}}
$$
By varying power radius locally, increasing power radius $r_j$ while fixing all others, the area of spherical power cell $R(G;c_i)$ increases monotonically, we find that $\frac{\partial{A_i}}{\partial{r_j}}\geq 0$. Thus Hessian matrix is diagonally dominated and semi-positive definite. By adding one more constraint such as fixing power radius at one point, the Hessian matrix becomes positive definite.
\subsection{Computational Method}
Based on above mentioned theories, we developed our specified algorithm for the lung nodules. In our new algorithm, all calculations, edge length, cell area, integration on a region, are under spherical metric. Exact formula for Hessian matrix is too complicated to present here. Nevertheless it's computable. Minimization of energy $E(\mathbf{r})$ can be computed by Newton's method, using the new formula of gradient and Hessian matrix. Figure~\ref{fig:OMT} shows a source sphere with measure and a target sphere with measure and the optimal transportation map.

\section{Experiments}\label{sec:Experiments}

The morphology of lung nodules have been shown to influence the likelihood ratios for malignancy ~\cite{Radiology1993,Acta1999}
and reflects the underlying pathological characteristics. In order to predict the likelihood ratios for malignancy of single lung nodule, we classified the nodules to malignant and benign, by using Wasserstein distance.

\subsection{Data Preparation}

The data of the lung nodule images were from computed tomography chest screens of patients visiting our institutions.
The interval between the slices were all around 1 mm.
All the nodules were surgically removed by the surgeons after diagnosis and pathologically verified for their malignancy.
The 3D models of the lung nodules were segmented and reconstructed by trained experts using commercial softwares and
confirmed by thoracic surgeons. Our experimental dataset had 55 lung nodules in total, including  39 malignant and  16 benign nodules,
and they were classified to different categories spiculation, lobulation, spherical by a thoracic surgeon
and a thoracic radiologist as the ground-truth. There are $23$ lobulation nodules, $17$ spiculation nodules, and $15$ spherical nodules in our dataset.

\subsection{Experiments Results}

For the classification purpose, we computed the full pair-wise Wasserstein distance matrix based on our new spherical optimal mass transportation algorithm. With the distance matrix, we validate our algorithm with lung nodule classification on a dataset of lung nodules, 39 malignant nodules and 16 benign nodules. The SVM method was employed as the classifier with 10-fold cross validation in our experimental results, and the input feature vector of the classifier includes 55 features, and the experiment result shows that our correctness rate is $80\%$.

The result presented in~\cite{Acta1999} shows that different morphology indicates different probability of the malignancy, There $82\%$ lobulation nodules, $97\%$ spiculation nodules were malignant and $66\%$ of spherical nodules were benign in their dataset. Our correctness rate match the average probability of malignancy/benign for nodules with different morphology, which indicates that our proposed algorithm analyze and classify the shapes correctly. However quantitative analysis of shape of the lung nodules may not enough to predict the malignancy in all cases.

\section{Conclusion}\label{sec:Conclusion}
\vspace{-3mm}
A novel 3D shape classification framework based on spherical
Wasserstein distance for lung nodules diagnosis is introduced in this paper.
We developed a specified optimal mass transport map on sphere for computing the Wasserstein distance for lung nodules.
Comparing to existing method, our algorithm is faster and more efficient. And our algorithm only depends on the Riemannian metric
and it is invariant under rigid motion and scaling.

We analyzed the 3D shapes of both malignant and benign nodule based on our proposed algorithm,
and were able to differentiate the malignant nodules from the benign ones. Our test results shows that using the Wasserstein distance
to describe the shape difference between lung nodules can constitute an efficient discriminatory feature, and it is very valuable in clinical practice, even not sufficient enough for all cases.

%

\bibliographystyle{unsrt}
\bibliography{miccai2018,IPMI_2015}

\end{document}